\documentclass[a4paper,11pt]{article}
\usepackage[left=2.5cm,right=2.5cm,top=2cm,bottom=2.0cm]{geometry}
\usepackage[ngerman, english]{babel}
\usepackage{color}
\usepackage{graphicx}
\usepackage{xspace}
\usepackage{amsmath}
\usepackage{lineno}
\usepackage{url}
\usepackage{hyperref}
\usepackage{bibentry}
\usepackage{bibgerm}

\title{Testing the EFT paradigm with top quark final states in proton-proton collisions at LHC with ATLAS and CMS}
\date{}

\author{Dennis Schwarz \\ on behalf of the ATLAS and CMS Collaborations}

\begin{document}
\newcommand{\pt}{\ensuremath{p_\text{T}}\xspace}
\newcommand{\ctqEight}{\ensuremath{C_{\text{tq}}^8}\xspace}
\newcommand{\ctG}{\ensuremath{C_{\text{tG}}}\xspace}
\newcommand{\ctZ}{\ensuremath{C_{\text{tZ}}}\xspace}
\newcommand{\ctZI}{\ensuremath{C_{\text{tZ}}^I}\xspace}
\newcommand{\ttbar}{\ensuremath{\text{t}\overline{\text{t}}}\xspace}
\newcommand{\ttbarjet}{\ensuremath{\text{t}\overline{\text{t}}}+jet\xspace}	
\newcommand{\ttgamma}{\ensuremath{\text{t}\overline{\text{t}}\gamma}\xspace}
\newcommand{\ttZ}{\ensuremath{\text{t}\overline{\text{t}}\text{Z}}\xspace}	
\maketitle


\begin{abstract}
	With the absence of the direct detection of new physics and the large data set collected at the LHC, the search for indirect effects in precision measurements became popular. Effective field theories enable to interpret these effects that originate from physics manifested at currently unreachable energy scales in a model independent framework. In this report, an overview of new analyses performed by the ATLAS and CMS Collaborations in the field of top quark physics is given.
\end{abstract}

\section{Introduction}
So far, the broad physics program at the Large Hadron Collider~(LHC) has not resulted in evidence for physics beyond the Standard Model~(BSM). Nonetheless, new physics may hide at energy scales beyond the reach of the LHC and thus, indirect searches may be able to reveal new particles or interactions at energy scales beyond the direct reach of the LHC. The experiments at the LHC have collected a vast amount of proton-proton collision data and with it precision measurements of known processes became feasible. Deviations in inclusive and differential cross sections can hint at new phenomena although they originate from much higher energies. Within the framework of effective field theories~(EFTs) these measurements can be interpreted in a model independent way. In current EFT interpretations, the Standard Model~(SM) Lagrangian is usually extended by including operators of dimension six, where Wilson coefficients~\footnote{All measurements reported in this article make use of the Warsaw basis~\cite{WarsawBasis} to define EFT operators and corresponding Wilson coefficients.} steer the coupling strength of the newly introduced operators. Due to their important role in the electroweak sector of the SM, top quarks are of special interest also in the field of EFT interpretations. In the following, recent highlights of EFT analyses connected to top quarks are presented, which were performed by the ATLAS and CMS Collaborations. Descriptions of the two experiments can be found in Refs.~\cite{ATLAS,CMS}.


\section{EFT interpretations from cross section measurements}
In general, two approaches of EFT interpretations exist at the ATLAS and CMS Collaborations. First, a cross section measurement can be reinterpreted by parameterizing the production cross section of a process in terms of Wilson coefficients. This approach allows to perform global fits, including also measurements that did not consider an EFT interpretation so far. A drawback of this approach is the definition of background processes. In a cross section measurement, all processes that are not considered as signal are subtracted from data. Thus, they are only considered at the SM point. However, these background processes can be affected by the EFT operators that are probed in the reinterpretation. Hence, the EFT effects in background processes are not taken into account. Especially in global fits this is problematic since this way one process may be varied to points in EFT parameter space in one analysis and fixed to the SM point in another.

\begin{figure}
	\centering
	\includegraphics[width=.49\textwidth]{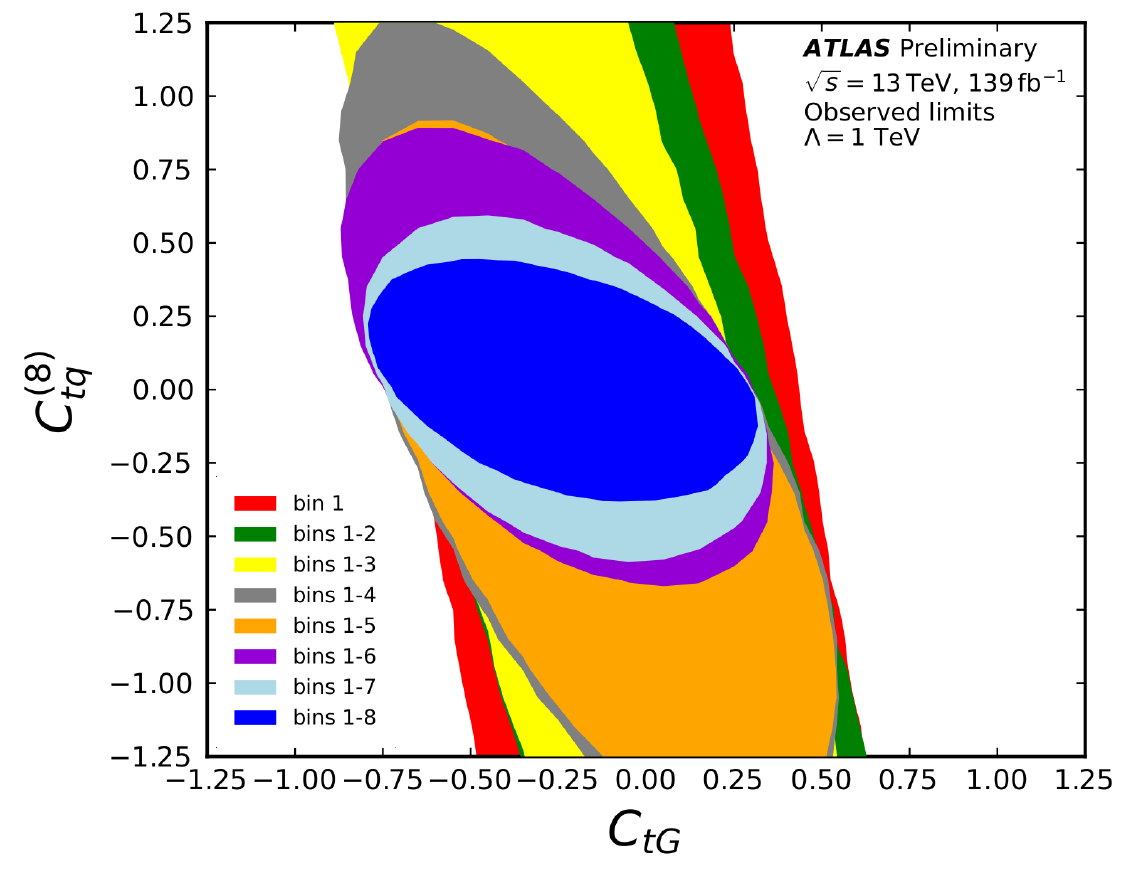}
	\includegraphics[width=.49\textwidth]{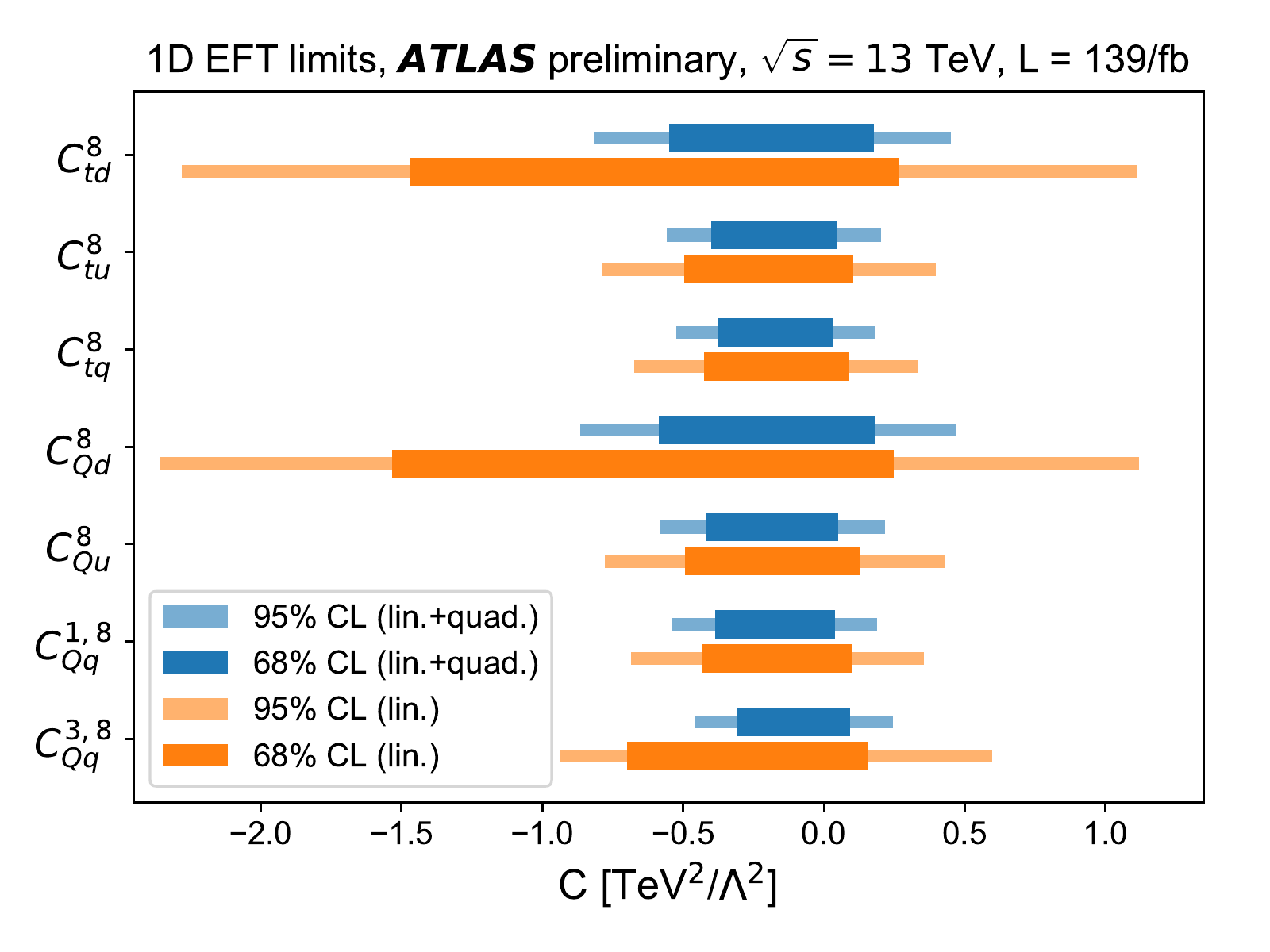}
	\caption{Evolution of two-dimensional bounds on \ctqEight and \ctG in the lepton+jets measurement when including bins with higher top quark \pt~(left), where bin 8 contains events with the largest top quark momenta. Summary of the resulting bounds on Wilson coefficients in the all-jets measurement~(right). Taken from Refs.~\cite{ATLASttboosted, ATLASttboostedhad}.}
	\label{f:boostedAllHad}
\end{figure}

A nonetheless well suited application for this approach is a measurement with a high signal purity. ATLAS performed measurements of the differential cross section as a function of various kinematic observables in the pair production of boosted top quarks \cite{ATLASttboosted, ATLASttboostedhad}. In both analyses, targeting the lepton+jets and all-jets channels of top quark-antiquark~(\ttbar) production, the cross section as a function of the hadronically decaying top quark transversal momentum (\pt) is parameterized in terms of Wilson coefficients. The interpretation of the differential cross section is sensitive to the inclusive \ttbar production on the one hand but also adds sensitivity to parameters that mostly affect events with high top quark momenta. In the lepton+jets measurement, the two coefficients \ctqEight and \ctG are tested. As Figure~\ref{f:boostedAllHad}~(left) shows, the bounds on \ctqEight get tighter when including events with larger top quark momenta while \ctG is already strongly constrained by the total cross section. In the measurement in the all-jets channel, the set of EFT parameters was chosen to contain only 2-light-quark 2-heavy quark coefficients, which are summarized in Figure~\ref{f:boostedAllHad}~(right). Both measurements feature a calibration of the jet energy scale using the reconstructed top quark mass. This procedure largely decreases the jet-related systematic uncertainties such that the all-jets measurement is already limited by statistical precision.

\begin{figure}
	\centering
	\includegraphics[width=.49\textwidth]{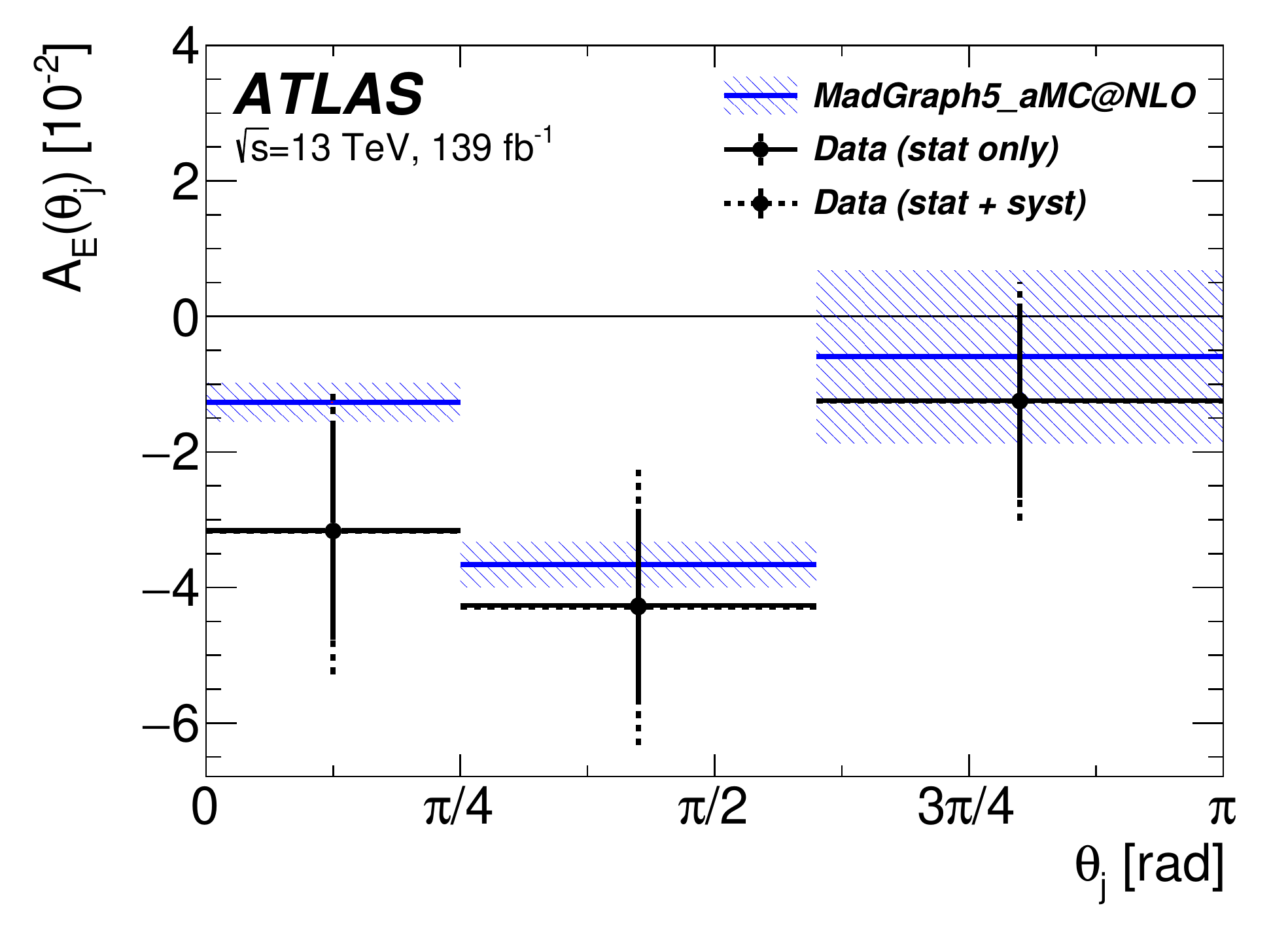}
	\includegraphics[width=.49\textwidth]{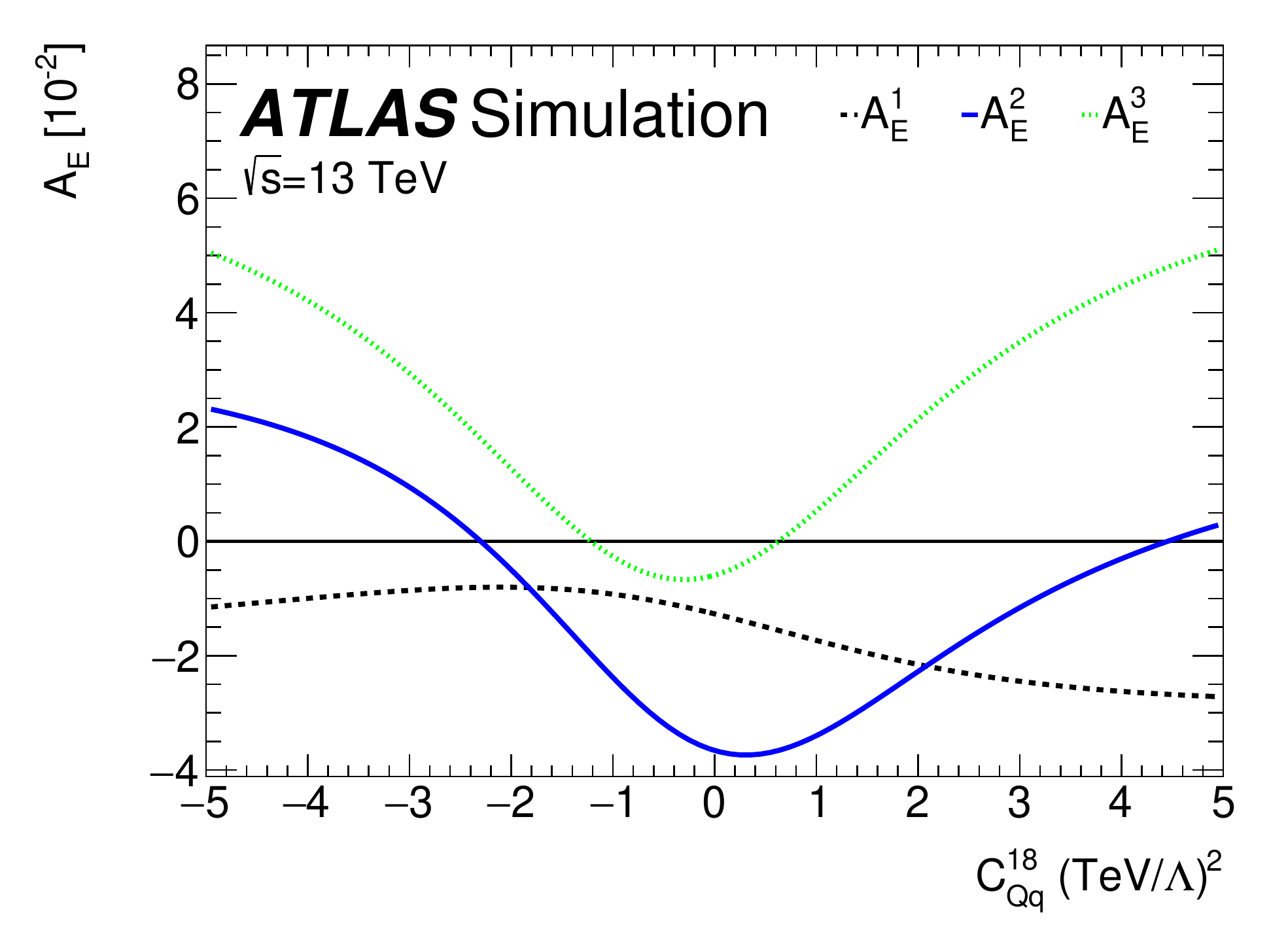}
	\caption{Top quark charge asymmetry~(left) in three bins of the angle of the additional jet to the beam axis and the sensitivity to one of the considered EFT parameters~(right). Taken from Ref.~\cite{ATLASttenergy}.}
	\label{f:ttenergy}
\end{figure}

The ATLAS Collaboration also published a measurement of the energy asymmetry of top quarks and top antiquarks in \ttbarjet production~\cite{ATLASttenergy}. Since this analysis aims at high-energy top quarks, it is complementary to the measurement of the top quark charge asymmetry~\cite{ATLASttcharge} that is dominated by threshold production. Figure~\ref{f:ttenergy} shows the energy asymmetry in three bins of the angle between the additional jet and the beam axis and the sensitivity to one of the EFT parameters considered. Also here the production cross section is written in terms of the Wilson coefficients of interest and then translated into the energy asymmetry. Thus, the result is reinterpreted to put bounds on EFT parameters. This measurement is limited by statistical precision. Important systematic uncertainties arise from the modeling of the signal in simulation and the jet calibration.

\section{Direct measurements of EFT parameters}
Recently, a second type of measurements became popular. Here, the analysis is already designed with the idea of an EFT interpretation. Often, multiple final states are addressed within one analysis and the bounds on Wilson coefficients are directly set from a sensitive distribution at the detector level. These specialized measurements offer the possibility to constrain a large number of EFT operators in a single analysis framework. In this scope, processes that are considered signal in one region and background in another as well as the correlations of systematic uncertainties can be handled in a consistent way.

\begin{figure}
	\centering
	\includegraphics[width=.42\textwidth]{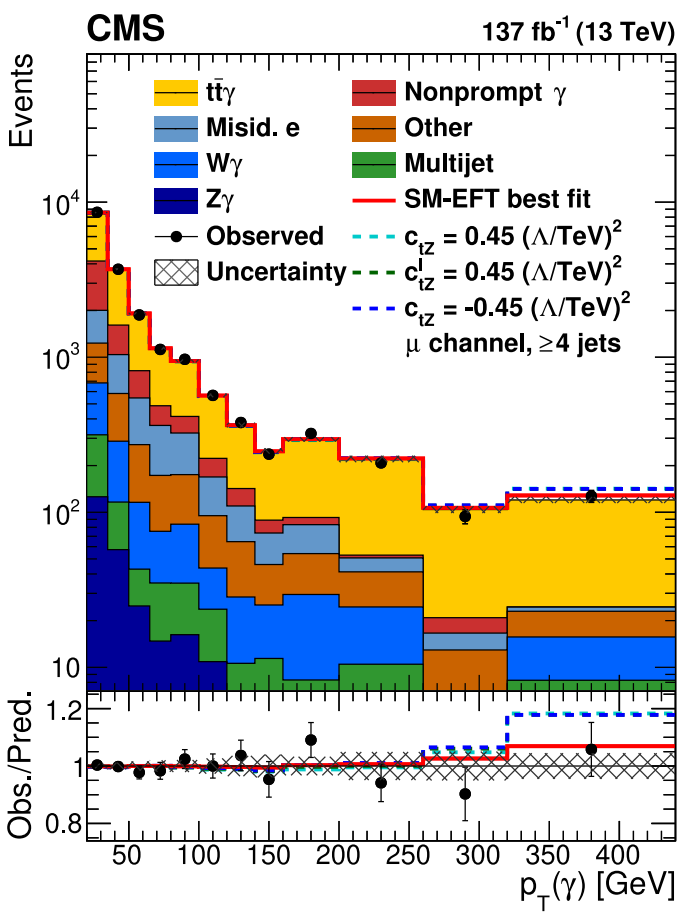}
	\includegraphics[width=.56\textwidth]{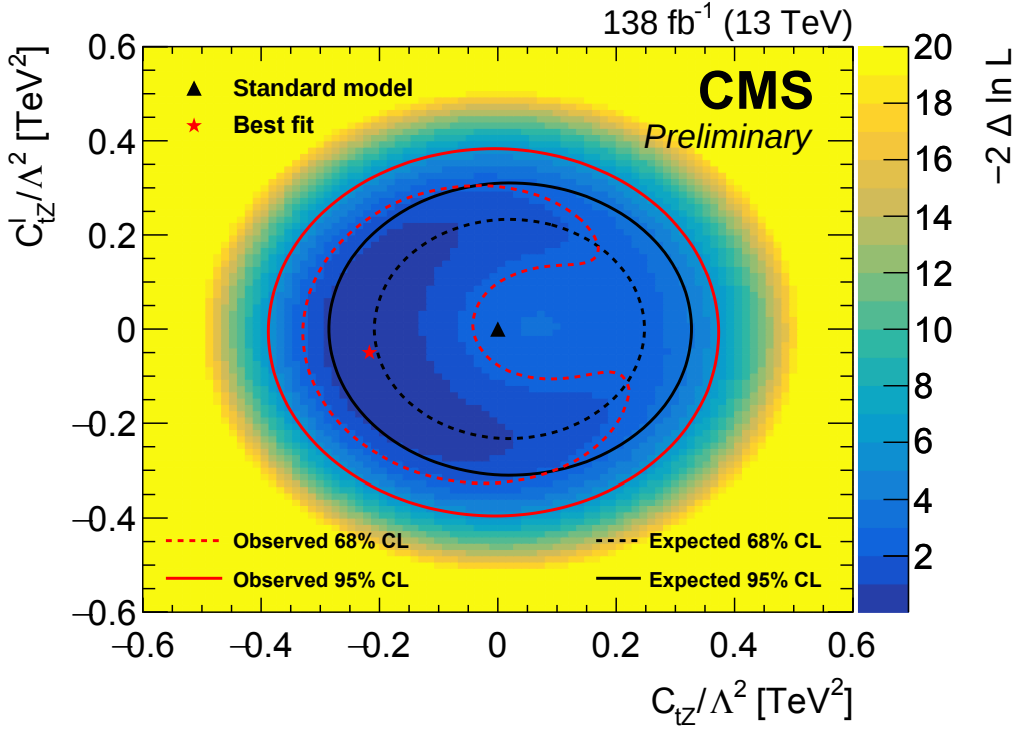}
	\caption{Photon \pt distribution in semi-leptonic channel of \ttgamma production (left) and combined limit with the dilepton channel (right). Taken from Refs.~\cite{CMSttgammasemilep, CMSttgammadilep}.}
	\label{f:ttgamma}
\end{figure}

The couplings of top quarks to electroweak gauge bosons are probed in top quark final states in association with photons or W/Z bosons. CMS published a measurement of \ttbar production with an additional photon in the final state. The measurement was performed in the semi-leptonic channel of \ttgamma~\cite{CMSttgammasemilep} and further combined with a measurement in the dilepton channel~\cite{CMSttgammadilep}. Figure~\ref{f:ttgamma}~(left) shows the photon \pt spectrum that was used to constrain the Wilson coefficients \ctZ and \ctZI in the semi-leptonic channel. Photons that originate from top quarks are likely to have higher momenta than those originating from the decay products of the top quark. Thus, the sensitivity to EFT operators that modify the top-photon coupling increase in the tails of the distribution. A crucial point in this measurement is the data driven estimation of background processes that contain misidentified photons and non-prompt leptons. The two-dimensional likelihood with 68\% and 95\% confidence intervals after the combination with the dilepton channel are displayed in Figure~\ref{f:ttgamma}~(right), which represent the tightest bounds on \ctZ and \ctZI so far. The sensitivity is driven by the semi-leptonic channel but can be slightly improved when combining with the dilepton channel. On the experimental side, the dominant uncertainties in this measurement are connected to the photon identification for the semi-leptonic measurement. The dilepton channel is limited by statistical precision. In both, large theoretical uncertainties are connected to the modeling of the signal process. In particular, the modeling of initial and final state radiation are the dominant uncertainties.

\begin{figure}
	\centering
	\includegraphics[width=.42\textwidth]{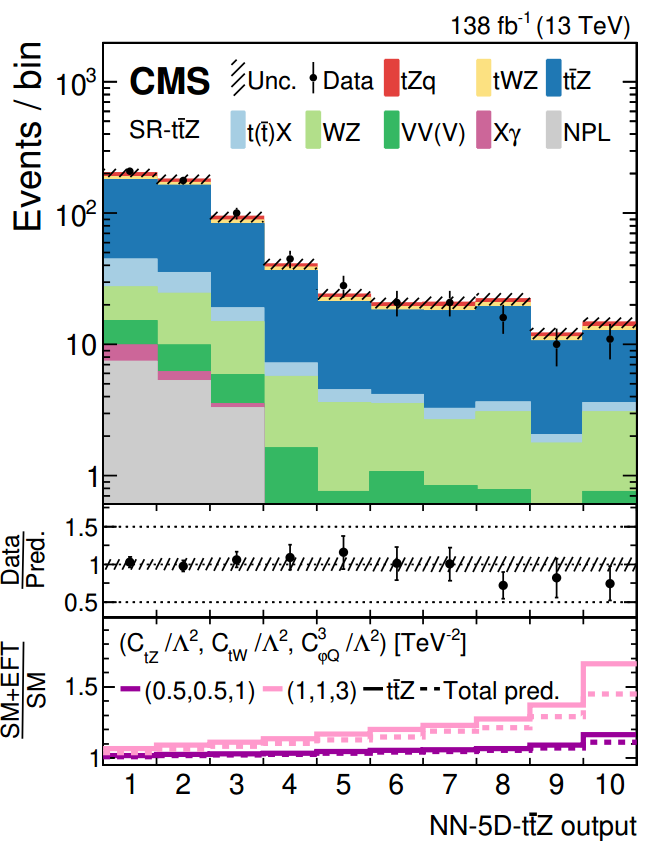}
	\includegraphics[width=.56\textwidth]{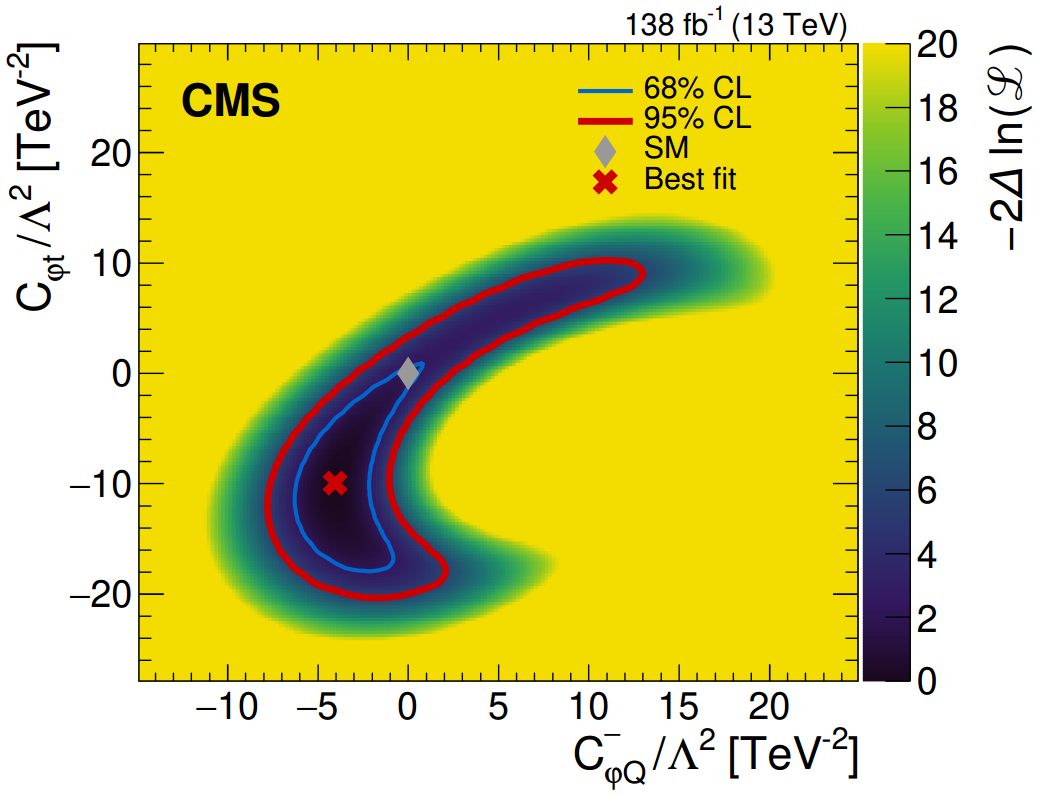}
	\caption{Output node of the classifier trained to detect events that are sensitive to variations of EFT parameters (left) and two-dimensional limits obtained from the measurement (right). Taken from Ref. \cite{CMSttML}.}
	\label{f:ttZ}
\end{figure}

Top quark production in association with Z and W bosons have been probed by both ATLAS~\cite{ATLASttV} and CMS~\cite{CMSttleptons, CMSttML}. The latest result by CMS~\cite{CMSttML} features an extensive use of machine learning techniques. One multiclass classifier is trained to separate the SM processes into three categories which are enriched in the tZq, \ttZ and background processes, respectively. A second set of networks is trained in the tZq and \ttZ regions to distinguish events with high sensitivity to variations in EFT parameter space from those with low sensitivity. Figure~\ref{f:ttZ} shows the output node of this second classifier in the \ttZ region. At higher values of the output, the sensitivity to EFT scenarios is highly increased. Limits are set by scanning single EFT operators, two at a time or even the full set of five operators that are considered in this analysis. The result of a two-dimensional scan is shown in Figure~\ref{f:ttZ}. Dominating uncertainties are connected to the background of non-prompt leptons on the experimental side.

\section{Conclusions}
The physics program of the ATLAS and CMS Collaborations already feature a broad range of EFT measurements in the top quark sector. Reinterpretations of cross section measurements enable a large compatibility to global fits, which combine several analyses of various fields in order to obtain a good overview on blank spaces in the EFT parameter space. On the other hand, dedicated measurements that are designed for the extraction of EFT parameters become more and more popular. With these, ambiguities in the treatment of backgrounds can be avoided and a large set of processes and couplings can be probed simultaneously. With the high-luminosity upgrade of the LHC, the future of EFT analyses looks even brighter since the increased data set will allow precision measurements of very rare electroweak processes.

\bibliographystyle{lucas_unsrt}
\bibliography{proceedings}

\end{document}